\documentclass{article}

\usepackage{amssymb}
\usepackage{hyperref}
\usepackage{amsmath}
\usepackage{graphicx}%
\usepackage{multirow}%
\usepackage{amsmath,amssymb,amsfonts}%
\usepackage{amsthm}%
\usepackage{mathrsfs}%
\usepackage[title]{appendix}%
\usepackage{xcolor}%
\usepackage{textcomp}%
\usepackage{manyfoot}%
\usepackage{booktabs}%
\usepackage{algorithm}%
\usepackage{algorithmicx}%
\usepackage{algpseudocode}%
\usepackage{listings}%
\usepackage{dsfont}
\usepackage{braket}
\usepackage{amssymb}
\usepackage{ragged2e}
\usepackage{hyperref}
\usepackage[utf8]{inputenc} 
\usepackage[T1]{fontenc}    
\usepackage{hyperref}       
\usepackage{url}            
\usepackage{booktabs}       
\usepackage{amsfonts}       
\usepackage{nicefrac}       
\usepackage{microtype}      
\usepackage{lipsum}
\usepackage{fancyhdr}       
\usepackage{graphicx}  
\newtheorem{theorem}{Theorem}
\newtheorem{proposition}[theorem]{Proposition}%

\theoremstyle{thmstyletwo}%

\theoremstyle{thmstylethree}%
\title{Quantum states as countable convex combination of pure states with bounded energy}

\author{
  Juan Pablo Lopez \\
  Universidad del Valle \\
  Cali, Colombia\\
  juan.lopez.holguin@correounivalle.edu.co \\
}

\begin{document}
\maketitle

\begin{abstract}
We give response to the question: in infinite dimension states,given a state with energy bounded by E, we can write the state as a countable convex combination of pure states with energy bounded by E. We review the Alicki-Fannes-Winter technique to obtain a uniform continuity bound for the von Neumann entropy in states that are a mix of pure states with bounded energy, using this bound we conclude that for a Hamiltonian satisfying the Gibb's hypothesis such states cannot exist.
\end{abstract}


\section{Introduction}

\indent In the paper \cite{Extreme} the authors proposed the question
\begin{center}
 {\textit{Under what conditions on the Hamiltonian H we could ensure that every state with energy bounded by E can be written as a countable convex combination of pure states with energy bounded by E. } }
\end{center}

\indent In the present paper we conclude that if H is a Hamiltonian satisfying the Gibb's hypothesis then we cannot express a state as an orthonormal countable convex combination of pure states with bounded energy. Our proof by contradiction is based in the construction of a uniform continuity bound of the von Neumann entropy and use this bound to conclude that such decomposition cannot be made it.\\
\indent The von Neumann entropy $S(\rho)=Tr(\rho log(\rho))$ is one of the most important measures in quantum information theory and is well know that in general is not continuous with the Schatten norm \cite{Shirkov Continuity}, however, in the last years some researchers have been developing continuity estimates which consist on finding functions $F$ and $g$ such that:

\begin{equation}
    |S(\rho)-S(\sigma)|\leq \epsilon F(\epsilon)+g(\epsilon),
\end{equation}
for all $\frac{1}{2}||\rho-\sigma||_1\leq \epsilon$. For the case that the density matrices have finite dimension we have the following bound:
\begin{equation}
    |S(\rho)-S(\sigma)|\leq \epsilon log_2(d-1)+g(\epsilon),
\end{equation}
with $g(\epsilon)=-\epsilon log_2(\epsilon)-(1-\epsilon)log_2(1-\epsilon)$ and d is the dimension of the Hilbert space that contains the density matrices, the sharpness of this bound was proved by Audenaert \cite{Audenaert}, meanwhile for the infinite dimensional case this bound becomes trivial and an energy constraint is necessary, Winter introduce in \cite{Winter} the bound
\begin{equation}\label{eq 3}
    |S(\rho)-S(\sigma)|\leq 2\epsilon S(\gamma (E/\epsilon))+g(\epsilon),
\end{equation}
and lately in \cite{Datta} was proposed an improvement, but an optimization problem is needed. In Sections \ref{sec 2} and \ref{sec 3} we review the Alicki-Fannes-Winter technique to apply it in \ref{sec 4} to states that are a mix of pure states with bounded energy to obtain a close form of uniform continuity for the entropy, the existence of this bound leads to a contradiction with the assumed Gibb's hypothesis.

\section{Gibbs hypothesis}\label{sec 2}
Consider a Hamiltonian H as an unbounded-Hermitian operator with discrete spectrum and bounded below, we consider the ground state of H as 0, we said that H satisfies the Gibbs hypothesis \cite{Winter} if for every $\beta>0$ we have $Tr(e^{\beta H})<\infty$. A $\textit{thermal state}$ has the form $e^{\beta H}/Z$ with $Z=Tr(e^{\beta H})$ is often the equilibrium state in most systems \cite{Thermal} and the unique maximizer of the von Neumann entropy $S(\rho)$ subject to the energy constraint $Tr(\rho R)\leq E$ is a thermal state denoted by $\gamma(E)$, if we consider $S(\gamma(E))$ as a function of $E\geq 0$ we see that is strictly increasing and concave \cite{Alicki}.Moreover, the von Neumann entropy is continuous in the set $\mathfrak{S}_{H,E}=\{\rho\in\mathfrak{T}^1: Tr(H\rho)\leq E\}$ \cite{Continuity} where $\mathfrak{T}^1 $ is the set of states. Finally, in the next section we present in detail the method used by Winter to obtain a continuity bound for the von Neumann entropy.

\section{Alicki-Fannes-Winter method}\label{sec 3}
One of the principal properties of the von Neumann entropy is the following form of concavity
\begin{equation}\label{eq 5}
    S(\lambda_1\rho+\lambda_2\sigma)\leq \lambda_1 S(\rho)+\lambda_2S(\sigma)+g(\lambda),
\end{equation}
where $h(\lambda)=-\lambda_1 log_2(\lambda_1)-\lambda_2 log_2(\lambda_2)$ and $\lambda_1+\lambda_2=1$, with this property Alicki and Fannes \cite{Alicki} obtained a continuity bound for quantum conditional entropy and von Neuman entropy for finite dimensional density matrices, later Winter \cite{Winter} uses the same strategy for obtain an equivalent bound in the infinite dimensional case and a Hamiltonian satisfying the Gibbs hypothesis. Now we review the Winter method,  consider the density matrices $\rho\in A_1$ and $\sigma\in A_2$ with spectral decomposition 
\[ \rho=\sum_i r_i \ket{e_i}\bra{e_i}  ,  \]
\[ \sigma=\sum_i s_i \ket{f_i}\bra{f_i}  , \]
where the eigenvalues are ordered in descending order. By the Mirksy's inequality we have
\begin{equation}\label{eq 5}
    2\epsilon:=\sum_i | r_i-s_i | =  ||(r_i)-(s_i)||_1\leq ||\rho-\sigma||_1.
\end{equation}
 We define the vector
\[\ket{\phi}:=\sum_i \sqrt{min\{r_i,s_i \}} \ket{e_i}^{A_1}\ket{f_i}^{A_2}, \]
and obtain the pure state $\ket{\phi}\bra{\phi}$ in $A_1A_2=A_1\otimes A_2$ 
\[ \ket{\phi}\bra{\phi}=\sum_i min\{r_i,s_i \} \ket{e_i}^{A_1}\ket{f_i}^{A_2} \bra{f_i}^{A_1}\bra{e_i}^{A_2} ,\]
where the trace is given by
\[ Tr(\ket{\phi}\bra{\phi})=\sum_i min\{r_i,s_i \}=\frac{1}{2}\sum_i (r_i+s_i-|r_i-s_i|)=1-\epsilon  ,\]
and the partial traces by
\begin{equation}
    \ket{\phi}\bra{\phi}^{A_1}=\sum_i min\{r_i,s_i \}\ket{e_i}\bra{e_i},
\end{equation}
\begin{equation}
    \ket{\phi}\bra{\phi}^{A_2}=\sum_i min\{r_i,s_i \}\ket{f_i}\bra{f_i}.
\end{equation}
 Finally we have that the operators
\begin{equation} \label{eq 9}
    \rho-\ket{\phi}\bra{\phi}^{A_1}=\sum_i (r_i-min\{r_i,s_i \}) \ket{e_i}\bra{e_i},
\end{equation}
\begin{equation} \label{eq 10}
    \sigma-\ket{\phi}\bra{\phi}^{A_2}=\sum_i (s_i-min\{r_i,s_i \}) \ket{f_i}\bra{f_i},
\end{equation}
are positive operators with trace equal to $\epsilon$, then we define the density matrices $\Delta_1=\epsilon^{-1}(\rho-\ket{\phi}\bra{\phi}^{A_1})$ and $\Delta_2=\epsilon^{-1}(\sigma-\ket{\phi}\bra{\phi}^{A_2}) $ to obtain the coupling $\omega=\ket{\phi}\bra{\phi}+\epsilon \Delta_1 \otimes \Delta_2$ of $\rho$ and $\sigma$. This decomposition allow us to use the \textit{strong subadditivity} and the concavity of the von Neumann entropy to obtain

\begin{equation}\label{Central}
    |S(\rho)-S(\sigma)|=|S(\omega^{A_1})-S(\omega^{A_2})|\leq S(\omega) \leq \epsilon S(\Delta_1\otimes \Delta_2)+h(\epsilon),
\end{equation}
where we need to bound the energy of $\Delta_1\otimes \Delta_2$ to apply the max-entropy principle, then we need to find the best possible bound for the energy of $\Delta_1\otimes \Delta_2$.

\section{ Mix of pure states with bounded energy }\label{sec 4}
\indent In \cite{Extreme} was proved that the extreme points of the convex set $\mathfrak{S}_{H,E}$ are the pure states with energy bounded by E, this set will be denoted by $\mathcal{H}_E=\{ \ket{\psi}\bra{\psi}:\bra{\psi}H\ket{\psi}\leq E \ , \ ||\psi||=1\}$. Let $\rho$ and $\sigma$ be states with the decomposition  

\[ \rho=\sum_i r_i \ket{e_i}\bra{e_i}  ,  \]
\[ \sigma=\sum_i s_i \ket{f_i}\bra{f_i}  , \]
where $ \bra{e_i}H\ket{e_i},\bra{f_i}H\ket{f_i} \leq E $, clearly $Tr(H\rho),Tr(H\sigma)\leq E$. In other words, $\rho$ and $\sigma$ are a \textit{countable convex combination} of pure states in $\mathcal{H}_E$. This states that are a mix of bounded energy pure states are special because we can make a great improvement in the von Neumann entropy, with this we can improve the  estimate of the energy of $\Delta_1\otimes \Delta_2$ as it is show in the following proposition.

\noindent \begin{proposition} \label{prop 1} Assume that $\rho$ and $\sigma$ are a \textit{countable convex combination} of pure states with bounded energy, then the energy of $\Delta_1\otimes \Delta_2$ is bounded by 2E.\\
\end{proposition}
\noindent\textbf{Proof: } The energy of $\Delta_1\otimes \Delta_2$ is simply $Tr(H\Delta_1)+Tr(\Delta_2)$, but using \eqref{eq 9} and \eqref{eq 10} we obtain that:
\[Tr(H\Delta_1)= \epsilon^{-1} \sum_i (r_i-min\{r_i,s_i \}) \bra{e_i} H \ket{e_i}, \]
\[Tr(H\Delta_2)= \epsilon^{-1} \sum_i (s_i-min\{r_i,s_i \}) \bra{f_i} H \ket{f_i} .\]
\indent For every $i\in\{1,2,...\} $  we have that  $\bra{e_i} H \ket{e_i},\bra{f_i} H \ket{f_i} \leq E$, then 
\begin{equation}\label{ec 11}
   Tr(H\Delta_1)\leq \frac{E}{\epsilon} \sum_i  (r_i-min\{r_i,s_i ) \},
\end{equation}
\begin{equation}\label{ec 12}
   Tr(H\Delta_2)\leq \frac{E}{\epsilon} \sum_i  (s_i-min\{r_i,s_i \}) .
\end{equation}
 \indent The energy of  $\Delta_1\otimes \Delta_2$ with the Hamiltonian $H^{A_1}\otimes \mathds{1}^{A_2}+\mathds{1}^{A_1}\otimes H^{A_2}$ is equal to $Tr(H\Delta_1)+Tr(H\Delta_2)$. Using (\ref{ec 11})  and (\ref{ec 12}) we obtain that 
\[  Tr(H\Delta_1)+Tr(H\Delta_2)\leq   \frac{E}{\epsilon} \sum_i  (r_i+s_i-2min\{r_i,s_i\}  )=\frac{E}{\epsilon} \sum_i |r_i-s_i|=2E,  \]
\begin{equation}
     Tr(H\Delta_1)+Tr(H\Delta_2)\leq 2E.
\end{equation}
\indent At this point Winter proposed $2E/\epsilon$ as bound for the sum of the energies of $\Delta_1$ and $\Delta_2$, the Proposition \ref{prop 1} makes an improvement of this bound and as consequence the following bound for the von Neumann entropy.\\
\noindent\begin{proposition} \label{theo 2} Let $\rho,\sigma$ be two density matrices with bounded energy $\\ Tr(H\rho),Tr(H\sigma)\leq E$ and $\frac{1}{2}||\rho-\sigma||_1= \epsilon\leq 1$, then we have 
\begin{equation}\label{eq 14}
    |S(\rho)-S(\sigma)|\leq 2\epsilon S\left(\gamma(E)\right) +h(\epsilon).
\end{equation}
\end{proposition}
\noindent\textbf{Proof: } Returning to the equation (\ref{Central}),  we use the Proposition \ref{prop 1} and maximum entropy principle to conclude the bound.\\ 

Compared to the previous bounds we see that \eqref{eq 14} do not have a dependency on $\epsilon$ in the function $S(\gamma(\cdot))$ which makes our bound to have be a closer uniform continuity bound. The fact that $\rho $ is decomposed in pure states with bounded energy implies the existence of an orthonormal basis $\{e_i\}_i$ with $\braket{He_i,e_i}\leq E$, now consider a state 

\[ \phi=\sum_i t_i \ket{\phi_i}\bra{\phi_i}  ,  \]
with $0<E \leq Tr(H\phi)\leq \tilde{E}$. We consider the state $\tilde{\phi}=\sum_i t_i \ket{e_i}\bra{e_i} $, we have that $S(\phi)=S(\tilde{\phi})$, this allow us to implement the bound (\ref{eq 14}) to every state with energy greater than E. 

\noindent\begin{proposition} \label{prop 3} Assume that there exists an orthonormal basis $\{e_i\}_i$ with $\\ \braket{He_i,e_i}\leq E$. Let $\rho,\sigma$ be two density matrices with  $E \leq Tr(H\rho),Tr(H\sigma)$ and $\epsilon$ defined as (\ref{eq 5}), then we have 
\begin{equation}\label{eq 15}
    |S(\rho)-S(\sigma)|\leq 2\epsilon S\left(\gamma(E)\right) +h(\epsilon).
\end{equation}
\end{proposition}
\noindent\textbf{Proof: } Let $\rho,\sigma$ have spectral decomposition
\[ \rho=\sum_i r_i \ket{\phi_i}\bra{\phi_i}  ,  \]
\[ \sigma=\sum_i s_i \ket{f_i}\bra{f_i}  . \]

Now if we consider the states $\tilde{\rho},\tilde{\sigma}$ defined as
\[ \tilde{\rho}=\sum_i r_i \ket{e_i}\bra{e_i}  ,  \]
\[ \tilde{\sigma}=\sum_i s_i \ket{e_i}\bra{e_i}  , \]
we have that $S(\rho)=S(\tilde{\rho})$ and $S(\sigma)=S(\tilde{\sigma})$, then 
\[ |S(\rho)-S(\sigma)|=|S(\tilde{\rho})-S(\tilde{\sigma} )|,  \]
with the states $\tilde{\rho},\tilde{\sigma}$ we can use the bound (\ref{eq 14}) to conclude the proof.\\
\indent Note that (\ref{eq 15}) if we ensure the existence of $\{e_i\}_i$ satisfying the condition $ \braket{He_i,e_i}\leq E$  we can conclude a bound that do not depend on the energy of the states and instead depends on a fixed energy E.

\begin{theorem}\label{Theo 4}
    Given a Hamiltonian H that satisfies the Gibb's hypothesis, then there is no orthonormal basis $\{e_i\}_i$ such that  $ \braket{He_i,e_i}=||\sqrt{H}e_i||\leq E$ for all i. Equivalently, we cannot express a state $\rho$ as an orthonormal countable convex combination of pure states with bounded energy.
\end{theorem}
\noindent\textbf{Proof: } Assume by contradiction that there exists an orthonormal basis $\{e_i\}_i$ such that  $ \braket{He_i,e_i}\leq E$ for all i. Consider a state $rho$ with energy $E\leq Tr(H\rho)$ and let $\sigma=\gamma(\tilde{E})$  with $E\leq \tilde{E}$, if we apply the Proposition \ref{prop 3} we have
\[  |S(\rho)-S(\gamma(\tilde{E}))|\leq 2\epsilon S\left(\gamma(E)\right) +h(\epsilon), \]

\[  -2S\left(\gamma(E)\right) -h(\epsilon)\leq S(\rho)-S(\gamma(\tilde{E}))  \leq 2S\left(\gamma(E)\right) +h(\epsilon), \]

\begin{equation}\label{key}
    S(\gamma(\tilde{E}))-2S\left(\gamma(E)\right) -h(\epsilon)\leq S(\rho)  \leq S(\gamma(\tilde{E}))+2S\left(\gamma(E)\right) +h(\epsilon),
\end{equation}
where $\frac{1}{2}||\rho-\sigma||_1\leq \epsilon$. If we fix $\rho$, as $ \tilde{E}$ increases we have that $S(\gamma(\tilde{E})) $ will increases until the left side of the inequality (\ref{key}) will be eventually greater than $S(\rho)$ giving a contradiction.

\section{Conclusions}
The von Neumann entropy is one of the most relevant quantities in information theory, it's applied to obtain bounds for maximal rate of information transfer \cite{Rate info} and other multiple information quantities. The search for uniform continuity bounds in the infinite dimensional case has been based in the Alicki-Fannes-Winter technique, our contribution is an application of this technique to certain states that are a mix of pure states with bounded energy to conclude that such states cannot exists, giving a partial response to the question made in \cite{Extreme}. Moreover,from the analysis we made for the proof we can conclude that in uniform continuity bounds we need a dependence on the energy that needs to be verified as we did in the Theorem \ref{Theo 4}.

\section*{Acknowledgments}
I would like to thank J. Delgado for his constant guidance. I am grateful to A. Winter for his commentary in the restriction of the states in the Proposition \ref{prop 1} .

\bibliographystyle{unsrt}

\end{document}